\documentclass[mathleft
]{an}
\usepackage{graphicx}
\usepackage{times}
\newcommand{\degg}{\hbox{$^\circ$}}

\newcommand{\xte}{{\it RXTE}}
\newcommand{\suzaku}{{\it Suzaku}}
\newcommand{\xmm}{{\it XMM-Newton}}
\newcommand{\chandra}{{\it Chandra}}
\newcommand{\sax}{{\it BeppoSAX}}
\newcommand{\ginga}{{\it Ginga}}
\newcommand{\asca}{{\it ASCA}}

\begin{document}

\Pagespan{789}{}
\Yearpublication{2006}%
\Yearsubmission{2005}%
\Month{11}%
\Volume{999}%
\Issue{88}%

\title{Suzaku Observations of Iron Lines and Reflection in AGN}

\author{J.N. Reeves\inst{1,2}\fnmsep\thanks{Corresponding author:
  \email{jnr@milkyway.gsfc.nasa.gov}\newline}
\and
A.C. Fabian\inst{3} \and
J. Kataoka\inst{4} \and
H. Kunieda\inst{5,6} \and
A. Markowitz\inst{1} \and
G. Miniutti\inst{3} \and
T. Okajima\inst{1,2} \and
P. Serlemitsos\inst{1} \and
T. Takahashi\inst{6} \and
Y. Terashima\inst{6,7} \and
T. Yaqoob\inst{1,2}
}
\titlerunning{Suzaku Observations of Iron Lines and Reflection}
\authorrunning{Reeves et al.}

\institute{Astrophysics Science Division, Code 662, NASA Goddard 
Space Flight Center, Greenbelt Road, Greenbelt, MD 20771, USA
\and 
Department of Physics and Astronomy, Johns Hopkins University, 
3400 N Charles Street, Baltimore, MD 21218, USA
\and 
Institute of Astronomy, University of Cambridge, Madingley Road, Cambridge, 
CB3 0HA, UK
\and
Department of Physics, Tokyo Institute of Technology, 2-12-1, Ohokayama, Meguro, 
Tokyo 152-8551, Japan
\and
Department of Physics, Nagoya University, Furo--cho, Chikusa, 
Nagoya 464-8602, Japan
\and
Institute of Space and Astronautical Science, Japan Aerospace 
Exploration Agency, Yoshinodai 3-1-1, Sagamihara, Kanagawa 229-8510, Japan
\and
Department of Physics, Ehime University, Matsuyama 790-8577, Japan}

\received{30 May 2005}
\accepted{11 Nov 2005}
\publonline{later}

\abstract{
Initial results on the iron K-shell line and reflection component in several AGN observed 
as part of the \suzaku\ Guaranteed time program are reviewed. This paper discusses 
a small sample of Compton-thin Seyferts observed to date with Suzaku; namely 
MCG\,-5-23-16, MCG\,-6-30-15, NGC\,4051, NGC\,3516, NGC\,2110, 3C\,120 and NGC\,2992. 
The broad iron K$\alpha$ emission line appears to be present in all but one of these 
Seyfert galaxies, while the narrow core of the line from distant matter 
is ubiquitous in all the observations. The iron line in MCG\,-6-30-15 shows the 
most extreme relativistic blurring of all the objects, 
the red-wing of the line requires the inner 
accretion disk to extend inwards to within $2.2R_{\rm g}$ of the black hole, 
in agreement with the \xmm\ observations. 
Strong excess emission in the Hard X-ray Detector (HXD) above 10\,keV 
is observed in many of these 
Seyfert galaxies, consistent with the presence of a reflection component from reprocessing 
in Compton-thick matter (e.g. the accretion disk). Only one Seyfert galaxy (NGC 2110) shows 
neither a broad iron line nor a reflection component. The spectral variability of 
MCG\,-6-30-15, MCG\,-5-23-16 and NGC 4051 is also discussed. In all 3 cases, the spectra 
appear harder when the source is fainter, while there is little variability of the iron line 
or reflection component with source flux. 
This agrees with a simple two component spectral model, 
whereby the variable emission is the primary power-law, while the iron line and 
reflection component remain relatively constant. }
\maketitle

\section{Introduction}

Determining the origin of the iron K emission line is one of the
fundamental issues in high energy research on AGN, as
it is one of the most direct probes available of the inner accretion
disk and black hole. The iron line  
first became important during the \ginga\ era, where observations
showed the 6.4 keV iron K$\alpha$ emission line was common amongst Seyfert
galaxies (Pounds et al. 1990). The associated 
``reflection hump'' above 10 keV, produced by Compton down-scattering 
of higher energy photons, showed the iron line emission arises from 
Compton-thick material, possibly the accretion disk (George \& Fabian 1991). 
The higher (CCD) resolution spectra available with \asca\ 
indicated that the iron line profiles were broad and asymmetrically skewed, 
which was interpreted as evidence
that the majority of the line emission
originated from the accretion disk around the massive black hole
(Tanaka et al. 1995, Nandra et al. 1997, Reynolds 1997). 


The picture emerging from \xmm\ and \chandra\ observations 
is more complex. The presence of 
a narrow 6.4 keV iron emission component, from more distant matter 
is common in many type I AGN (Yaqoob \& Padmanabhan 2004, Page et al. 2004). 
Furthermore, complex absorption in some objects
could effect the modeling of the iron K-line and reduce its strength 
(Reeves et al. 2004, Turner et al. 2005). 
The presence of a reflection component hardening the spectrum towards higher 
energies (George \& Fabian 1991) also complicates fitting the broad iron 
K line, if bandpass above 10\,keV is not available to determine its strength.

New observations with the \suzaku\ X-ray observatory can provide an important 
breakthrough in this area, by determining the true underlying AGN continuum 
emission over a wide bandpass (e.g. from 0.3 keV to $>100$\,keV), 
thereby resolving the ambiguities present in fitting the iron K-shell band. 
Importantly the bandpass of above 10\,keV provided by the hard X-ray detector (HXD) 
on-board \suzaku\ makes it possible to measure the reflection component 
simultaneously with the iron K line. 
In the next section the \suzaku\ observatory is briefly 
outlined, 
while in Section 3 results from individual observations of Seyfert galaxies performed 
as part of the \suzaku\ Guaranteed Time (GT) program are outlined. The majority of
the subsequent spectra 
show detections of the relativistic iron line and associated Compton reflection 
hump.

\begin{figure}
\begin{center}
\includegraphics[width=80mm,height=60mm,angle=0]{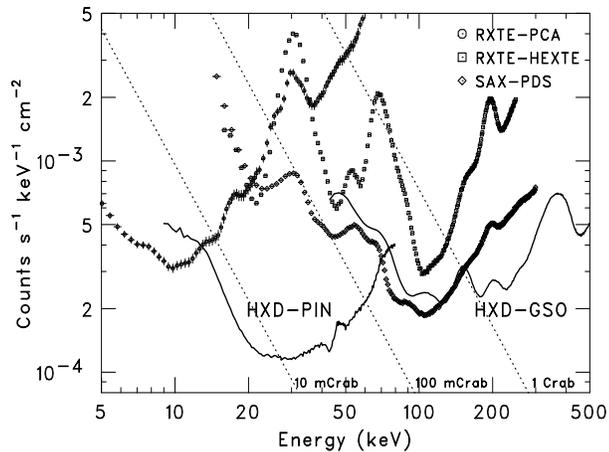}
\caption{The \suzaku\ HXD non X-ray background compared to other missions 
(Mitsuda et al. 2006). The background level 
for each instrument is normalized to its effective area (in cm$^{2}$). The HXD PIN and 
GSO background is shown as a solid line over the 10-500\,keV band, for comparison the relative 
background level for \sax\ PDS  and \xte\ PCA/HEXTE are plotted, along 
with the flux as a proportion of the Crab (dotted lines). 
The background level for the HXD is low, e.g. $<10$\,mCrab at 20 keV.}
\end{center}
\end{figure}

\section{The Suzaku X-ray Observatory}

Astro-E2 (subsequently renamed \suzaku), the fifth in the series of Japanese X-ray 
astronomy satellites, was launched from Uchinoura Space Center, Japan on July 10, 2005. 
It has reached a low Earth orbit (570\,km altitude, with an inclination of 31\degg), with 
an orbital period of 96 minutes (Mitsuda et al. 2006). 
The working scientific payload of \suzaku\ 
consists of four X-ray focusing telescopes (Serlemitsos et al. 2006), with four X-ray sensitive 
CCDs at the focal plane of the telescopes. These CCDs (or X-ray Imaging 
Spectrometers, XIS, Koyama et al. 2006) consist of three front illuminated (FI) 
CCDs (sensitive from 0.4--12\,keV) and one back illuminated CCD (BI) CCD (sensitive from 
0.2--12\,keV) with enhanced soft X-ray response. The FI chips are also referred to as 
XIS\,0, XIS\,2 and XIS\,3 respectively, while the BI CCD is referred to as XIS\,1. 
The second instrument on-board \suzaku\ is a non-imaging, collimated Hard X-ray Detector 
(HXD, Takahashi et al. 2006), which extends the \suzaku\ bandpass out to 
higher energies from 10--600\,keV. The HXD consists of three detectors, the PIN (silicon diodes 
sensitive from 10--70\,keV), the GSO (a phoswich counter 
sensitive from 40--600\,keV), as well as the non-pointing 
HXD-WAM (Wide-band All-sky Monitor).

In this paper, we discuss the data only from the XIS, HXD/PIN and HXD/GSO. The 4 XIS 
detectors have a combined collecting area similar to XMM-Newton; 
$\sim1000$\,cm$^{2}$ at 6 keV, while at 1.5\,keV 
the XIS area is $\sim1300$\,cm$^{2}$.  
The energy resolution of the XIS is 130\,eV (FWHM) at 6 keV. 
The low energy line spread function of the XIS is nearly symmetrical in shape, 
allowing the detection of low energy K-shell lines from C, N and O.  
At hard X-ray energies above 10\,keV, the HXD has high sensitivity and low background 
compared to other missions. 
A comparison of the \suzaku\ HXD non X-ray background versus Beppo-SAX 
PDS and RXTE PCA/HEXTE is shown in Figure\,1. The HXD 
background level is lower compared to the PDS instrument over most of the energy range; 
the HXD PIN background is $<10$ mCrab at 20 keV. The
systematic uncertainty of the HXD background is currently 5\% (with a 1\% goal), 
making it possible to study hard X-ray faint astronomical sources 
($<1$\,mCrab). Note especially the low background level of the HXD at the peak energy 
of the Compton reflection hump in AGN or of the Cosmic hard X-ray 
background (near 20--30\,keV). 
Also, given the AGN being detected in the Swift 
BAT All Sky Survey above 15 keV (Markwardt et al. 2005), it will be possible to perform a 
broad band spectral study of $>200$ AGN at a flux level $>1$\,mCrab with both XIS and HXD. 
These properties of \suzaku\ are therefore of great potential benefit to the study of 
the iron line and the Compton reflection component in AGN.

The \suzaku\ observations described in this paper were processed with the latest version 
of the pipeline available to the Science Working Group (SWG) members. Typical screening 
parameters used to process and extract the data are described elsewhere; e.g. 
refer to Reeves et al. (2006) for a description of the AGN MCG\,-5-23-16, or to Terada et al. 
(2006) for the HXD analysis of the binary X-ray pulsar, AO\,0535+262. 
For the XIS, the soft X-ray contamination (in the C and O band) has been corrected 
in the spectra.   
For the HXD/PIN, background spectra were extracted from a time-dependent model provided by 
the instrument team. Cosmic X-ray background (CXB) flux 
(Gruber et al. 1999) was also included in the 
PIN background model, at a level of 
$9.0\times10^{-12}$\,erg\,cm$^{-2}$\,s$^{-1}$, integrated over the 
15--50\,keV band. The HXD/GSO data are only discussed here in the observations of the 
two brightest AGN in this sample; in MCG\,-5-23-16 and NGC\,2110 the GSO background 
was extracted from observations of the clusters Abell\,1795 and Fornax 
respectively. 

\begin{figure}
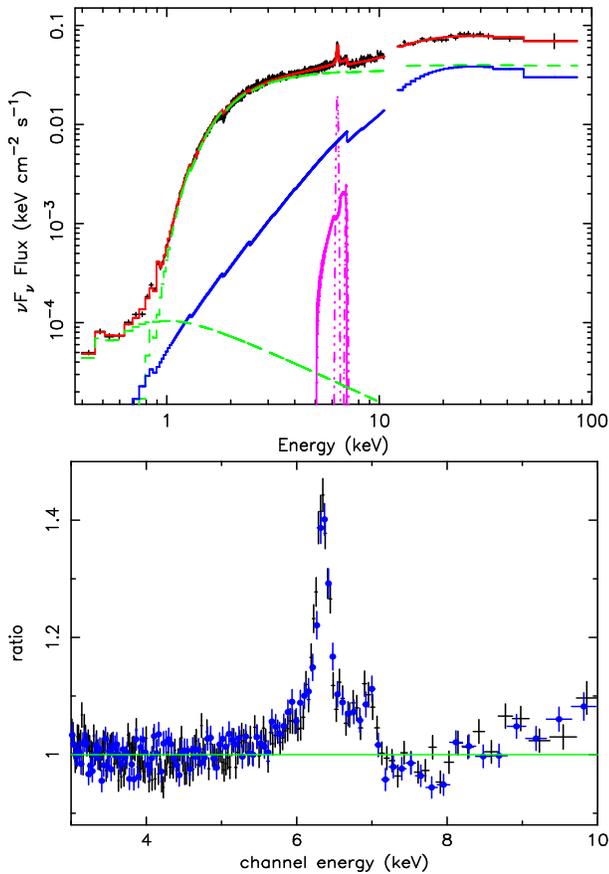

\begin{center}
\includegraphics[width=60mm,height=80mm,angle=-90]{eeuf.ps}
\includegraphics[width=55mm,height=80mm,angle=-90]{mcg5_fe.ps}
\caption{The upper panel shows the \suzaku\ spectrum of the Seyfert 1.9 galaxy, MCG\,-5-23-16, 
The black data points represent the XIS and HXD data, the solid red line the total 
model, the dashed green lines the direct (absorbed) hard X-ray and  
scattered soft X-ray (unabsorbed) power-law continua. The iron K line emission is 
represented by the magenta lines, the solid line shows emission from the diskline and 
the dot--dashed lines are narrow iron K$\alpha$ and K$\beta$ lines. Notice the 
presence of a reflection hump above 10\,keV, the reflection 
model being represented by the solid blue line. The lower panel shows 
the ratio of the iron line profile in MCG\,-5-23-16 to the best-fit continuum model, 
as observed simultaneously 
with \suzaku\ XIS FI (black crosses) and \xmm\ EPIC-pn (blue circles). The observations 
are 100\,ks in length. Both a broad red-wing and 
a narrow core to the line are clearly visible.}
\end{center}
\end{figure}

\section{X-ray Spectra of Seyfert Galaxies Obtained with Suzaku}
\subsection{MCG\,-5-23-16}

The Compton-thin Seyfert 1.9 galaxy MCG -5-23-16 (at $z=0.008486$) was observed by 
\suzaku\ between 7--10 December 2005, for a total exposure time of 98.1\,ks with XIS. 
The \suzaku\ observation was performed simultaneously 
with \xmm\ and \chandra. 
The broad band spectral energy distribution from XIS and HXD  
is shown in Figure\,2. The spectrum is plotted with respect to a baseline continuum 
model consisting of a power-law of $\Gamma=1.93\pm0.03$, absorbed by neutral matter of 
column density $N_{\rm H}=1.65\pm0.03\times10^{22}$\,cm$^{-2}$. A steep 
soft power-law (of $\Gamma\sim3$) 
attenuated only the the local Galactic column density of $8\times10^{20}$\,cm$^{-2}$ is 
also included in the continuum model. This is likely to arise due to indirect scattered 
X-ray continuum as well as photoionized emission from distant gas. 
The spectrum is then well fitted ($\chi^{2}/{\rm dof}=1983/1879$) when a Compton 
reflection component with $R=1.1\pm0.2$ is added (here $R=\Omega/2\pi$, where $\Omega$ 
is the solid angle subtended by the reflecting material), while the iron abundance is 
$0.6\pm0.1$ times Solar. Overall, the source flux of MCG\,-5-23-16 
was high, $8.8\times10^{-11}$\,erg\,cm$^{-2}$\,s$^{-1}$ in the 2-10 keV band or 
$1.9\times10^{-10}$\,erg\,cm$^{-2}$\,s$^{-1}$ from 15-100\,keV.

Figure\,2 
also shows a close-up of the iron K line profile for MCG\,-5-23-16, notice the 
excellent agreement between the XMM-Newton EPIC-pn data and Suzaku XIS. The narrow 
iron K$\alpha$ line core at 6.40\,keV as well as the K$\beta$ 
line are both clearly detected, the narrow K$\alpha$ core has an equivalent width of 70\,eV 
and is unresolved in both \suzaku\ XIS and \chandra\ ($<4600$\,km\,s$^{-1}$ FWHM), 
consistent with an origin in distant matter far from the black hole. 
Nonetheless a broad iron K line is unambiguously detected, with a similar equivalent width 
as per the narrow core (60 eV), broadened by a velocity width of $\sigma\sim0.1c$. 
When the broad component is modeled by a disk line around a Schwarzschild black hole 
(Fabian et al. 1989), the inner disk radius is $37^{+20}_{-10}R_{\rm g}$, 
with a disk inclination angle of  $\sim50\degg$. More details of the spectral 
fitting are given in Reeves et al. (2006). An inclination of 50\degg\ 
is consistent with the classification of MCG\,-5-23-16 as a Seyfert 1.9 galaxy within 
AGN unified schemes.

\begin{figure}
\begin{center}
\includegraphics[width=55mm,height=80mm,angle=-90]{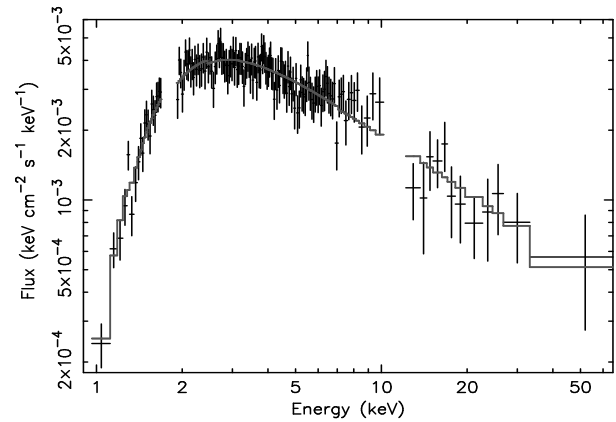}
\caption{The \suzaku\ difference spectrum of MCG\,-5-23-16, fitted with a simple 
absorbed power-law of $\Gamma=1.9$. The difference spectrum shows that the iron line and 
reflection hump appear to be less variable than the power-law continuum.}
\end{center}
\end{figure}

The spectral variability of AGN can also be explored with \suzaku, over a 
broad energy band and especially in the reflection spectrum above 10\,keV. 
Splitting the MCG\,-5-23-16 observation into high and low flux spectra 
revealed no apparent change 
in the iron line flux or absolute strength of the reflection component. The difference 
spectrum of the high -- low flux states from 1--50\,keV is shown in Figure\,3, 
which is consistent with a 
simple absorbed $\Gamma=1.9$ power-law continuum with no excess due to an iron K line nor a 
reflection hump above 10\,keV. Thus it appears that the variable emission component is 
the primary $\Gamma=1.9$ power-law, while the reflection spectrum remains roughly constant.
In the MCG\,-5-23-16 observation, the continuum is variable by 40\%, while any changes 
in the reflection are limited to $<20$\%. The lack of variability of the reflector may 
in part be due to the distant reflector, note the relative strengths of the  
broad and narrow iron lines suggest they contribute equally to the reflection spectrum.

\begin{figure}
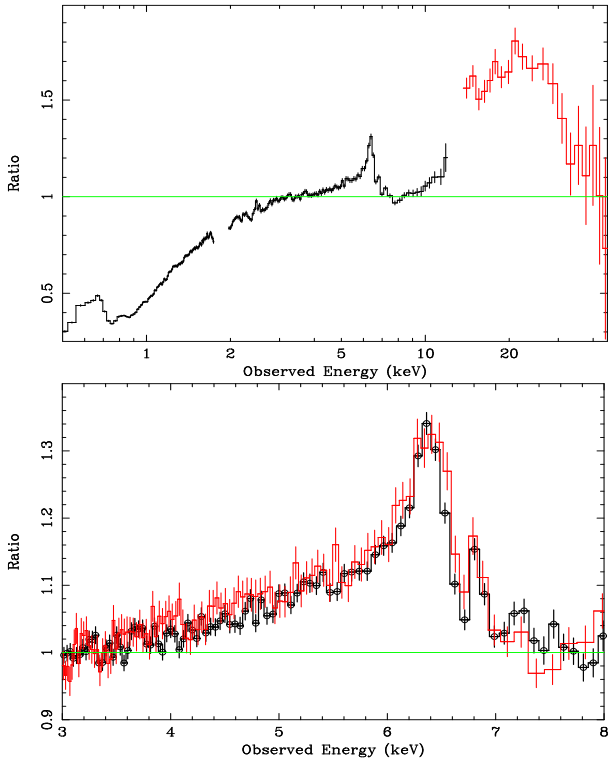

\begin{center}
\includegraphics[width=50mm,height=80mm,angle=-90]{mcg6_ratio1.ps}
\includegraphics[width=50mm,height=80mm,angle=-90]{mcg6_compare.ps}
\caption{Spectra from the deep 350\,ks \suzaku\ observation of MCG\,-6-30-15. The upper plot 
shows the broad band Suzaku spectrum from XIS FI and HXD/PIN, plotted as a ratio to 
a $\Gamma=2.0$ power-law, fitted over the 3.0--4.0\,keV and 7.5--10\,keV bands. 
The effect of the warm absorber is seen below 3\,keV, while the relativistic 
iron line profile is observed between 4--7\,keV. The HXD/PIN data above 10\,keV reveals 
the unambiguous detection of a strong reflection hump associated with the broad iron line, 
with $R>2$. The lower panel shows a close-up of the iron line profile, comparing the 
deep Suzaku observation with the previous deep \xmm\ exposure. 
The Suzaku XIS points are shown as black open circles, the XMM-Newton EPIC-pn as 
red crosses. Notice the structure 
in the blue-wing of the line, which is due to resonance iron K-shell absorption 
at 6.7\,keV and 7.0\,keV.}
\end{center}
\end{figure}

\subsection{MCG\,-6-30-15}

The Seyfert 1 galaxy MCG\,-6-30-15 ($z=0.00775$) was observed 3 times by \suzaku\ 
in 2006 January, with a total 
exposure of 347\,ks and a mean 2-10\,keV flux of 
$4.0\times10^{-11}$\,erg\,cm$^{-2}$\,s$^{-1}$, similar to the 
\xmm\ deep look (Fabian et al. 2002). The broad band \suzaku\ spectrum of MCG\,-6-30-15 
plotted as a ratio to a power-law continuum of $\Gamma=2.0$, is shown in Figure 4. The 
familiar signature of the broad iron K line is present between 4--7\,keV, while below 
3\,keV the spectrum is strongly absorbed by a warm absorber as previously 
known (e.g. Turner et al. 2004). Importantly the 
HXD/PIN data above 10\,keV show a very strong hard excess, the clear signature of a  
reflection component. The iron line profile is also shown in Figure 4 (lower panel), 
which shows a remarkable similarity to the profile obtained with the 
deep \xmm\ observation (Fabian et al. 2002; Vaughan \& Fabian 2004). 

\begin{figure}
\begin{center}
\includegraphics[width=50mm,height=80mm,angle=-90]{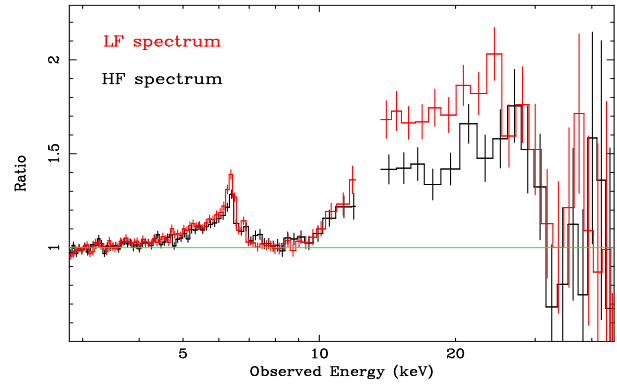}
\caption{High and low flux spectra of MCG\,-6-30-15, plotted against a power-law continuum. 
The low flux spectrum shows a stronger reflection component when compared to the 
weaker power-law continuum level. This implies that the reflection component 
is not varying along with the continuum.}
\end{center}
\label{label1}
\end{figure}

As shown in Miniutti et al. (2006), the iron line can be parameterized with a double 
Gaussian to represent the red-wing 
and blue-core of the line, with a total equivalent width of 
$320\pm45$\,eV. The blue core of the line near 6.4\,keV is clearly resolved with 
$\sigma>260$\,eV (implying emission within $<80R_{\rm g}$ of the black hole), while the 
red-wing has a width of $\sigma>760$\,eV, implying 
emission within $<6.5R_{\rm g}$ of the black hole. The unresolved iron line core emission is 
constrained to only $30\pm5$\,eV in equivalent width, 
Furthermore, two weak absorption lines from He and H-like iron are observed near 6.7\,keV 
and 7.0\,keV within the blue-wing of the line, as also observed by \chandra\ HETG (Young et al. 
2005). As the Fe K-shell absorption is highly ionized (and crucially 
no absorption from species below He-like Fe are detected), this suggests the absorber has 
a negligible effect on the relativistic line in the iron K band. 

The spectrum above 3 keV can be well fitted (reduced $\chi^{2}=1.08$, for 2312 
degres of freedom) with a blurred ionized disk reflection model
(Ross \& Fabian 2005). The inner disk radius in constrained to within 
$R_{\rm in}<2.2R_{\rm g}$ of the black hole (the outer radius is fixed at $400R_{\rm g}$), 
which places an lower-limit on the black hole spin of $a>0.917$ (i.e. near maximal 
for a Kerr metric). The disk inclination 
is found to be $38\pm4$\degg. The disk emissivity can be parameterized by a broken
power-law function, with a steep inner emissivity of $q=4.6^{+0.6}_{-0.9}$, a break 
radius fixed at $6R_{\rm g}$ and a flatter outer emissivity of $q=2.6\pm0.3$. 
(Note the emissivity $q$ is parameterized as $r^{-q}$, where r is the radius from the 
black hole). 
The reflection component is very strong, with $R=2.8\pm0.9$ 
representing $51\pm10$\% of the continuum flux from 14-45\,keV, 
while the ionization state is close to neutral ($\xi=65\pm45$\,erg\,cm\,s$^{-1}$) 
and the iron abundance is 2 times solar. 
As the narrow unresolved core of the line is 
so weak, the reflection is dominated by the inner disk, the distant reflector having 
a negligible contribution. The reflection spectrum appears to be enhanced with respect to 
the power-law continuum, which 
could argue in favour of gravitational light bending of continuum photons towards 
the innermost disk (Miniutti \& Fabian 2004).

It is also possible to examine the spectral variability 
over a wide energy band with XIS and HXD and therefore measure the variability of the 
reflection component directly. As per the earlier XMM-Newton observations, MCG\,-6-30-15 shows 
short term continuum variability of a factor of $\times2-3$ over a few ks, making it possible 
to derive high and low flux spectra for the source. Figure 5 shows the high and low flux 
spectra, plotted against a power-law continuum. The reflection hump appears stronger 
in the low flux spectrum against the weaker continuum ($R_{\rm low}=4.8\pm0.8$ vs. 
$R_{\rm high}=2.5\pm0.6$), consistent with there being no change in the absolute 
strength of the reflection component. This is confirmed by the high -- low 
difference spectrum, which shows no variable component to the Compton hump above 10\,keV. 
Furthermore a flux--flux analysis, plotting the 14--45\,keV band HXD flux points 
against the 1--2\,keV band XIS points (Miniutti et al. 2006), 
shows a constant hard offset component equal 
in magnitude to the reflection component. Thus it appears that the disk 
reflector is less variable than the power-law continuum, which is predicted 
by the light bending model of Miniutti \& Fabian (2004).

\begin{figure}
\begin{center}
\includegraphics[width=80mm,height=60mm]{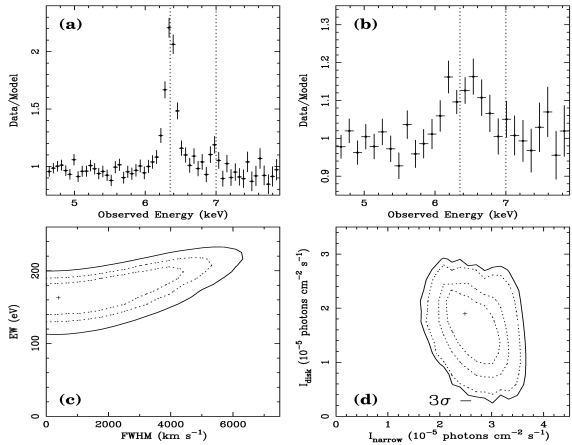}
\caption{The iron line profile of NGC 2992 measured by \suzaku. Panel (a) shows the 
clear detection of the narrow K$\alpha$ and K$\beta$ lines. Panel (b) plots the iron 
line profile after the K$\alpha$ and K$\beta$ lines have been fitted, showing a clear broad 
base with velocity broadening of about 0.1c. Panel (c) shows the 68, 90 and 99\% 
contours of the narrow K$\alpha$ line width versus its equivalent width. Panel (d) shows the 
69, 90 and 99\% and 3$\sigma$ contours of the narrow versus broad line K$\alpha$ line 
intensity, which demonstrates that the broad and narrow lines are decoupled at $>3\sigma$ 
confidence.}
\end{center}
\end{figure}

\subsection{NGC 2992}

The Compton-thin Seyfert 1.9 NGC\,2992 ($z=0.00771$) 
was observed 3 times by \suzaku\ in 2005, November and December, with a total 
exposure time of 108\,ks. The source was found to be in a relatively low state, 
with a 0.5--10\,keV unabsorbed luminosity of $2.5\times10^{42}$\,erg\,s$^{-1}$.
The XIS data was analyzed in detail by Yaqoob et al. (2006), while the HXD data was not 
presented at 
this stage, due to the low flux level of the source. One of the main results 
was that the broad and narrow iron line could be decoupled at a high degree 
of confidence. This is illustrated in Figure 6. 
Most importantly panel (d) in Figure 6 shows the confidence contours of the 
narrow versus broad iron line flux, showing that the two components are decoupled at 
$>3\sigma$ confidence. The equivalent width of the broad line was found to be 
$118^{+32}_{-61}$\,eV. The most likely origin of the line is the accretion disk. 
For an assumed disk emissivity of $q=3$, the line can be characterized by an inner disk 
radius of $\sim31R_{g}$ and an inclination angle $>29$\degg. 
Note although a Compton down-scattered shoulder to the 6.4\,keV line 
core would be expected near 6.24\,keV (e.g. Matt et al. 2002), its intensity and 
width are negligable compared to the broad iron line.  

\begin{figure}
\begin{center}
\includegraphics[width=80mm,height=30mm]{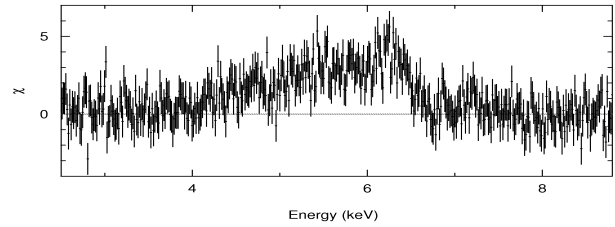}
\caption{
The \suzaku\ broad iron line profile of NGC 3516 plotted as residuals 
against the best fit continuum model, after the warm absorber, reflection 
component and the narrow iron line core have been modelled.}
\end{center}
\end{figure}

\subsection{NGC 3516}

The Seyfert 1 NGC 3516 was observed by Suzaku in 2006, October 12--15, for a total 
of 135\,ks exposure. The findings are briefly summarized here, they are 
presented in more detail by Markowitz et al. (2006). 
The 2--10\,keV flux observed by \suzaku\ was similar to the 2001 \xmm\ 
observations (Turner et al. 2005) of about $3.0\times10^{-11}$\,erg\,cm$^{-2}$\,s$^{-1}$; 
however in the soft band (0.5-2.0\,keV) the flux measured by \suzaku\ is a factor of 
$\times2-3$ lower than in \xmm. 
A detailed analysis of the 2001 \xmm\ and \chandra\ HETG 
data by Turner et al. (2005) suggested that a broad red-wing to the iron line was not 
required, while the spectrum could be modeled by multiple layers of a complex absorber, 
producing spectral curvature in the 2--6\,keV band. In the Suzaku observation, 
a broad relativistic iron K line is present (Figure 7), with an 
equivalent width of $185^{+130}_{-70}$\,eV 
and can be parameterized by disk emission down to a 
radius of $<5R_{\rm g}$. The line core is also resolved, with a FWHM of 
$3800^{+1500}_{-1900}$\,km\,s$^{-1}$, consistent with the expected width from the 
BLR in this object (Peterson et al. 2004). 
Both the broad iron line and the complex absorption are required in the \suzaku\ data, 
with an increase in the column density of a 
factor $\times3$ between the \xmm\ and \suzaku\ observations, consistent with the decrease 
in soft flux. A reflection component is detected in HXD/PIN consistent with $R=1$ covering, 
while the underlying power-law continuum is well constrained, with $\Gamma=1.84\pm0.02$. 
This demonstrates the ability of Suzaku to deconvolve a complex spectrum over a broad 
bandpass, helping break the degeneracy between the continuum, absorption, reflection and 
a complex iron line profile. 

\begin{figure}
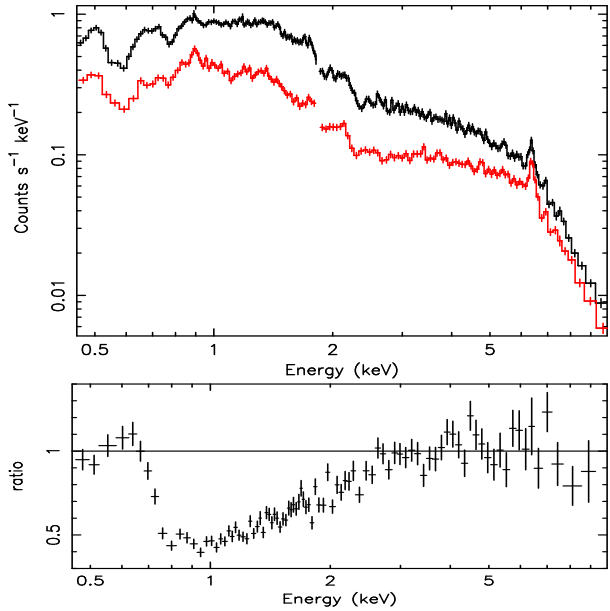

\begin{center}
\includegraphics[width=50mm,height=80mm,angle=-90]{ngc4051_hilow.ps}
\includegraphics[width=30mm,height=80mm,angle=-90]{ngc4051_diff.ps}
\caption{The top panel shows XIS (FI) high and low flux spectra for NGC 4051. 
Overall NGC\,4051 was observed at a low flux level, compared to the historical mean. 
The source clearly appears to be harder when fainter. The lower panel shows the 
high -- low difference 
spectrum as a ratio to a steep $\Gamma=2.1$ power-law. No variable component of the 
iron line appears to be present, while the soft band is modified by a warm absorber 
between 0.7--3\,keV.}
\end{center}
\end{figure}

\subsection{NGC 4051}

NGC\,4051, a Narrow Line Seyfert 1 at $z=0.002336$ was observed by \suzaku\ for 
a net exposure of 93.2\,ks. The AGN shows typical large amplitude 
fluctuations making it possible to extract high and low flux spectra 
from the \suzaku\ observation. These spectra are shown in Figure\,8, 
showing that NGC 4051 is harder at low fluxes, and less variable 
in the hard band. The source has a flat spectrum in the 2-10\,keV band, with 
$\Gamma=1.55$ and $\Gamma=1.15$ for the high and low spectra, with fluxes of 
$1.1\times10^{-11}$\,erg\,cm$^{-2}$\,s$^{-1}$ and $6.3\times10^{-12}$\,erg\,cm$^{-2}$\,s$^{-1}$
respectively. Both fluxes are lower than the historical mean for this AGN, 
the low flux spectrum is comparable to the low-state \xmm\ observation in 2002
($5.8\times10^{-12}$\,erg\,cm$^{-2}$\,s$^{-1}$; Pounds et al. 2004). The high -- low 
difference spectrum shows that the variable emission is represented by a much steeper 
$\Gamma=2.1\pm0.1$ power-law, modified by a warm absorber between 0.7--3\,keV, as is shown 
in Figure\,8. No iron line emission 
appears to be present near 6\,keV in the difference spectrum 
indicating that the iron line is relatively constant, 
with the steep power-law being the intrinsically variable component. 

Figure 9 shows the mean spectrum as a ratio against this intrinsic $\Gamma=2.1$ 
continuum. It appears that absorption has a substantial effect on the spectrum below 
about 3--4\,keV, producing considerable spectral curvature. At the softest energies an excess 
of emission is present, as well as emission in photoionized gas (the 0.57\,keV O\,VII 
line is readily apparent). Above the iron line a strong hard excess is apparent, likely  
associated with a reflection component. 

\begin{figure}
\begin{center}
\includegraphics[width=55mm,height=80mm,angle=-90]{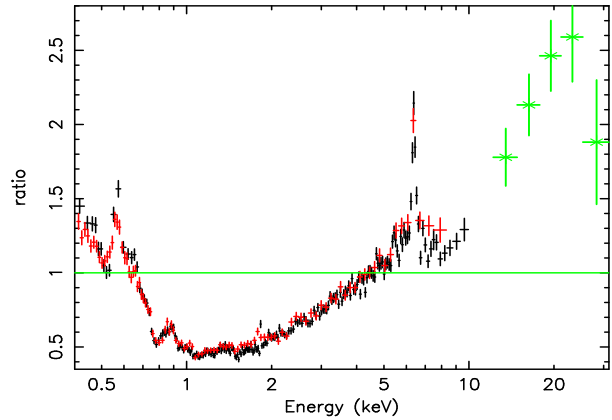}
\caption{The mean spectrum of NGC\,4051, plotted as a ratio against the intrinsically 
variable $\Gamma=2.1$ continuum. The XIS FI is shown in black, XIS BI in red and 
HXD/PIN in green. The effect of complex absorption below 4\,keV is clear, while 
a hard excess appears to be present above the iron line.}
\end{center}
\end{figure}

Indeed the spectrum can be well modeled by including complex/ionized absorption, 
a soft excess including
soft X-ray line emission, broad and narrow iron K lines and Compton reflection, 
superimposed on the variable power-law of $\Gamma=2$. Posssible origins for the 
soft X-ray lines include emission from distant photoionized gas (Pounds et al. 2004) or 
from the BLR or even the accretion disk (Ogle et al. 2004).
The iron emission line is 
strong, the unresolved narrow core having an equivalent width of $135\pm20$\,eV, 
and the broad line $250\pm60$\,eV (against the direct continuum). 
The broad line is well fitted by a profile from a 
disk extending inwards to $1.2R_{\rm g}$, with an emissivity profile of $q=3$. The reflection 
component appears strong, with $R>2$. The strength of the iron line and 
reflection may be due to the low continuum level of NGC\,4051, if 
indeed the line/reflection component is not rapidly responding to the continuum flux 
(also see Ponti et al. 2006). 
The \suzaku\ spectrum suggests that not only is 
complex absorption required, as favoured by Pounds et al. (2004), but a broad emission 
line and an inner disk reflector are also present.

\subsection{3C 120}

The broad line radio galaxy 3C\,120 (at $z=0.033$) was observed 4 times by \suzaku\ 
during 2006 in February and March, with a total net exposure of 145\,ks with XIS. 
The spectrum can be well modeled with an absorbed power-law of $\Gamma=1.84\pm0.02$, 
a reflection component at high energies consistent with $R=0.5\pm0.2$ and both
narrow and broad components of the iron 
line. The narrow line core is resolved, with a velocity 
width of 8200\,km\,s$^{-1}$ ($\sigma=76\pm11$\,eV), which is broader than the typical 
uncertainties in the XIS response function width at 6\,keV.  This 
is consistent with the line width in the previous \xmm\ observation 
(Ballantyne et al. 2004, Ogle et al. 2005). The line core could arise from the BLR or 
from the outer parts of the disk at $\sim10^{3}R_{\rm g}$. 
However there is also a highly broadened 
component to the line, as is evident in Figure 10, which was not so apparent in the 
\xmm\ observation. The broad component is 
centered near 6.4\,keV and has a FWHM of 62000\,km\,s$^{-1}$ (or $\sigma=580\pm110$\,eV). 
The line can be well modeled by a disk extending inwards to $\sim20R_{\rm g}$. 

\begin{figure}
\begin{center}
\includegraphics[width=55mm,height=80mm,angle=-90]{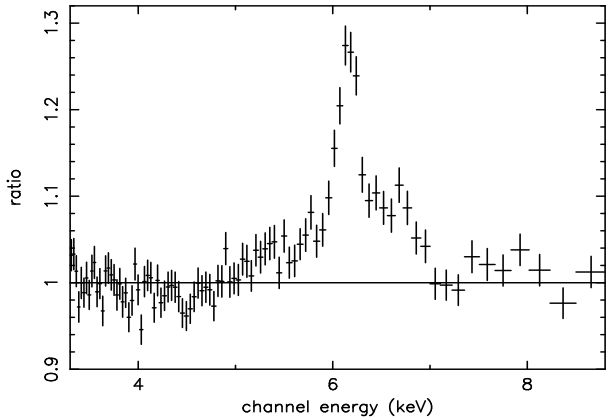}
\caption{The iron line profile of 3C\,120, as measured by the Suzaku XIS, compared 
to an absorbed power-law of $\Gamma=1.84$. The line profile shows both broad and 
narrow components.}
\end{center}
\end{figure}

\subsection{NGC 2110}

The Compton-thin Seyfert 2 NGC 2110 (at z=0.0078) was observed by Suzaku in September 
2005, for a net exposure of 100\,ks (Figure 11) and at a very high flux level, of 
$1.05\times10^{-10}$\,erg\,cm$^{-2}$\,s$^{-1}$ (2-10\,keV) or 
$2.7\times10^{-10}$\,erg\,cm$^{-2}$\,s$^{-1}$ (15-100\,keV). A weak unresolved 
iron line was detected by \suzaku, at $6.40\pm0.01$\,keV, with an equivalent width of 
$44\pm3$\,eV. The stringent limit on the reflection component with HXD of $R<0.1$ 
implies an origin in Compton-thin matter, while the high energy 
cut-off is restricted to $>200$\,keV. No diskline component 
appears to be present in this source, down to an upper-limit of 10\,eV (assuming 
diskline parameters of 6.40\,keV rest frame energy, 
$R_{\rm in}=6R_{\rm g}$, emissivity $q=-3$ and inclination 45\degg). 
Thus NGC\,2110 appears to be the only AGN here which does not show a broad / relativistic 
iron line, or show evidence of reflection. 

\begin{figure}
\begin{center}
\includegraphics[width=55mm,height=80mm,angle=-90]{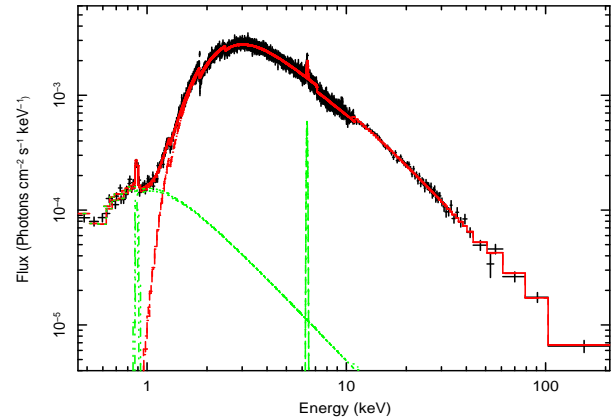}
\caption{The broad band spectrum of the Compton-thin Seyfert 2, NGC 2110, as 
observed by Suzaku XIS, HXD/PIN and HXD/GSO. The iron line is narrow, while the 
complete lack of any reflection ($R<0.1$), suggests an origin in transmission through 
Compton-thin matter. The continuum is absorbed at low energies, through a neutral 
column of $3\times10^{22}$\,cm$^{-2}$. A scattered soft X-ray component (green line) 
is also observed below 1 keV.}
\end{center}
\end{figure}

\section{Discussion and Conclusions}

The Suzaku observations confirm the existence 
of the relativistic iron line in 6 out of the 7 Seyfert galaxies. All of the 
AGN with broad iron K lines have detections of strong Compton reflection 
components in the HXD data, consistent with the broad line originating in 
Compton-thick matter such as the accretion disk. The only Seyfert galaxy without a broad iron 
line is NGC\,2110, which also does not appear to show any evidence of reflection, down 
to a tight limit of $R<0.1$. Importantly the broad band-pass of Suzaku makes it possible 
to resolve any degeneracies in modelling the iron K line profile. 
By accurately accounting for both 
the soft X-ray absorption and the reflection component, 
it is possible to derive the primary underlying continuum emission, which is crucial 
for modeling the broad red-wing of the iron line. While complex 
ionized absorption can have a substantial effect on the soft X-ray continuum, e.g. as 
discussed in NGC\,3516, NGC\,4051 and MCG\,-6-30-15, the broad iron K line is required as well. 
Importantly the reflection components detected in several of the AGN (e.g. MCG\,-6-30-15) 
appear too strong to originate purely from distant matter (such as the torus), as the 
narrow iron K$\alpha$ line cores are relatively weak, a few tens of eV in 
equivalent width. 

As would be expected given the previous observations with \asca\ (Tanaka et al. 1995) 
or XMM-Newton (Fabian et al. 2002), MCG\,-6-30-15 shows the most extreme 
relativistic iron line profile, requiring the inner radius of the accretion disk 
to extend to within $<2.2R_{\rm g}$ of the black hole. 
Such a robust upper-limit to the innermost stable orbit of the disk constrains the spin of 
the black hole to $a>0.917$, implying near maximal rotation.
The fact that the reflection component in MCG\,-6-30-15 is so strong ($R>2$), could 
suggest that the effects of General Relativity (GR) near the innermost stable orbit are 
important 
in the X-ray spectrum. For instance the light-bending model as proposed by 
Miniutti \& Fabian (2004) would suppress the amount of continuum observed, 
if the continuum photons are bent away from the observer and towards the disk/black 
hole plane, thereby enhancing the relative strength of the reflection component. 

The broad lines in NGC\,3516 and NGC\,4051 are consistent with disk emission  
extending inwards to the innermost stable orbit, i.e. at least $6R_{\rm g}$ in the 
Schwarzschild case or 
even $1.2R_{\rm g}$ for a Kerr metric. Although broad lines are required in MCG\,-5-23-16, 
NGC\,2992 and 3C\,120, the lines (typically 0.1c in FWHM) 
could be characterized by emission from a few tens of gravitational radii. 

The presence of the narrow iron K$\alpha$ line core appears to be ubiquitous, being detected in 
all of the spectra presented here. In some cases, the line core appears unresolved, which 
would suggest reprocessing in distant matter, such as the torus 
(Ghisellini, Haardt \& Matt 1994), an outflow (Elvis 2000), or 
a bi-conical structure (Sulentic et al. 1998). In at least two cases (3C\,120 and NGC\,3516)  
the iron K$\alpha$ line core is resolved, with a velocities consistent 
with the AGN BLR. In the example of NGC\,2110, the narrow line must 
originate in transmission from Compton-thin matter, 
due to the lack of the reflection hump, while in MCG\,-5-23-16 
the reflection component is strong and the unresolved 
line core is more likely to originate from scattering off distant Compton-thick matter. 
Thus while the detection of narrow Fe lines appears ubiquitous, the 
physical origin of the line may differ, or have contributions from different 
regions.

One of the strengths of Suzaku is that it is possible to study the spectral variability 
of AGN over a broad bandpass, enabling both the variability of the iron 
line and the reflection component above 10\,keV to be measured. In the three examples 
discussed here, namely MCG\,-6-30-15, MCG\,-5-23-16 and NGC 4051, it appears that the 
iron line is less 
variable than the continuum. The unique aspect of Suzaku 
shows that the reflection hump is less variable than the continuum. 
For MCG\,-6-30-15, where a substantial 
proportion of the X-rays originate from radii closest to the black hole event horizon, the 
effect of GR could be important in dampening the amplitude of variability in the reflection 
component, as per the light bending model. Alternatively, in MCG -5-23-16 or NGC\,4051, where 
the narrow cores of the iron K$\alpha$ line are stronger, then 
the presence of a constant distant reflector could dilute any possible variability 
in the line or reflection hump. While the current observations do suggest that the iron line 
and reflected emission is not responding to the X-ray continuum level on short timescales, 
this does not 
exclude the possibility of short-term variations which are not correlated with 
continuum flux. For instance, short-lived, transient redshifted 
iron emission line features have been detected in several AGN (e.g. Turner et al. 2002; 
Iwasawa, Miniutti \& Fabian 2005), albeit at modest statistical significances. 
Thus it will be important to 
monitor variations in the iron line and reflection component in time as well as 
with continuum flux. 



\acknowledgements
We would like to thank the entire Suzaku Science Working Group for their contribution 
towards the mission. In particular the hard work of all the instrument teams is 
acknowledged, who have made all these observations possible.



\end{document}